\documentclass[12pt,preprint]{aastex}
\begin{document}
\slugcomment{astro-ph/0404434 CGPG--04/4--5}
\shorttitle{Universality of the subsolar mass distribution...}
\shortauthors{Visser and Yunes}

\title{\bf Universality of the subsolar mass distribution \\
from critical gravitational collapse}

\author{Matt Visser}
\affil{School of Mathematical and Computing Sciences, \\
Victoria University of Wellington, \\ 
P.O. Box 600, Wellington,  New Zealand}
\email{matt.visser@mcs.vuw.ac.nz}
\and
\author{Nicolas Yunes}
\affil{Center for Gravitational Physics and Geometry, \\
Center for Gravitational Wave Physics, \\
Department of Physics, \\
The Pennsylvania State University,\\
University Park, Pennsylvania 16802, USA}
\email{yunes@gravity.psu.edu}

\date{\centerline{19 April 2004; \LaTeX-ed \today}}
%
%
\def\d{{\mathrm{d}}}
\def\implies{\Rightarrow}
\def\be{\begin{equation}}
\def\bd{\begin{equation}}
\def\ba{\begin{eqnarray}}
\def\bea{\begin{eqnarray}}
\def\ee{\end{equation}}
\def\ed{\end{equation}}
\def\ea{\end{eqnarray}}
\def\eea{\end{eqnarray}}
\def\ie{{\emph{i.e.}}}
\def\eg{{\emph{e.g.}}}
\def\p{{\partial}}
\def\N{I\!\!N}

\begin{abstract}
  Self-similarity induced by critical gravitational collapse is used
  as a paradigm to probe the mass distribution of subsolar objects.
  At large mass (solar mass and above) there is widespread agreement
  as to both the form and parameter values arising in the mass
  distribution of stellar objects. At subsolar mass there is still
  considerable disagreement as to the qualitative form of the mass
  distribution, let alone the specific parameter values characterizing
  that distribution.
 
  For the first time, the paradigm of critical gravitational collapse
  is applied to several concrete astrophysical scenarios to derive
  robust qualitative features of the subsolar mass distribution. We
  further contrast these theoretically derived ideas with the
  observational situation.  In particular, we demonstrate that at very
  low mass the distribution is given by a power law, with an exponent
  opposite in sign to that observed in the high-mass regime. The value
  of this low-mass exponent is in principle calculable via dynamical
  systems theory applied to gravitational collapse.  Qualitative
  agreement between theory, numerical experiments, and observational
  data is good, though quantitative issues remain troublesome.
\end{abstract}


\section{Theoretical background}

The initial mass function [IMF] describing the mass distribution of
stellar objects is one of the basic tools used in studying the
evolution and lifecycles of stars and galaxies.  At large mass, the
IMF is well-characterized by a power law with the Salpeter exponent
$1.35$.  At small sub-solar mass the situation is much more uncertain.
Using theoretical ideas based on self-similarity and critical
gravitational collapse, we argue in a model-independent manner that
there \emph{must} be a change in this power law, and in fact that the
sign of the exponent must flip.  Contrasting this with direct
observation indicates that the IMF is certainly modified below
approximately $0.8 \; M_{\odot}$, and the theoretical expectation of a
sign flip is borne out by the data. [The current article fleshes out
some of the details backing up the comments briefly summarized
in~ \citet{Yunes04}.]

Our key observation starts from the well-known fact that gravitational
condensation, either Newtonian or general relativistic, is
characterized by the existence of critical exponents and power-law
behavior.  By linearizing around any critical solution at the
threshold of collapse, the mass $M$ of the resulting condensed object
is related to any suitable control parameter $A$ in the initial data
by an equation of the
form~\citep{Harada03,Harada01,Gundlach1,Gundlach2,Gundlach3,Gundlach4}
\be 
M \approx M_0 \; [A-A_{\mathrm{critical}}]^\delta;
\qquad \delta > 0.  
\ee 
The fact that scaling of this type often occurs is \emph{generic}, and
independent of the details of the collapse process and equation of
state --- and independent of whether or not the collapse is
relativistic. The precise \emph{value} of the scaling exponent does
depend on the specific details of the collapse process, but the fact
that such scaling occurs is intrinsically universal.

Once a scaling law of this type is derived, we shall see that
straightforward manipulations lead to a power law for the distribution
of low-mass objects
\be
P(M) \approx \frac{\cal A}{M_0} \; 
\left(\frac{M}{M_0}\right)^{(1/\delta)-1},
\ee
with an exponent that is calculable in terms of the mass-scaling
exponent, $\delta$. In this manner, we can explain the low-mass tail
in the Initial Mass Function [IMF] from first principles in terms of
dynamical systems theory in gravitational collapse. The technique
developed in this article cannot say anything about the high-mass tail
of the IMF, but that is a regime where there is reasonable theoretical
and observational agreement on the state of affairs. We shall
specifically concentrate on the functional form of the IMF for
subsolar masses.
 
To set the stage, recall that any gravitationally self-interacting
cloud of gas, either Newtonian or general relativistic, has a limited
number of long-term fates:
\begin{itemize}
\item The cloud can completely disperse to infinity.
  
\item Part of the cloud might condense, with the remainder dispersing
  to infinity.
  
\item The entire cloud might condense.
\end{itemize}
The condensed object could, for instance, be a solid planet, a fluid
star, or a black hole, although it does not need to collapse to
infinite density. The set of all initial data that lead to any one of
these fates can be thought of as an infinite-dimensional phase space,
containing infinite-dimensional basins of attraction for each final
fate. Since there are three possible final fates for a cloud of gas,
there will be three basins of attraction: the collapse basin, where
its attractor leads to complete collapse; the dispersal basin, for
which the final fate is an asymptotically flat Minkowski spacetime;
and an intermediate collapse basin, where ultimately part of the cloud
collapses and the rest disperses to infinity. These basins will be
separated from each other by boundaries of co-dimension one, or
separatrices, that form the so-called critical surfaces. In this
manner, it is clear that the critical surfaces contain all critical
initial data that separate two basins of attraction.  An example of
critical initial data, {\it{i.e.}} a point on the critical surface,
would be the Jeans mass, or Jeans energy.  Another important point on
this surface will be an intermediate attractor in phase space, which
will be referred to as the critical solution. This critical solution
will have important properties, such as self-similarity or
scale-invariance.  For a more complete and detailed analysis refer
to~\citep{Gundlach1,Gundlach2,Gundlach3,Gundlach4}.

Applying dynamical systems theory to a region of phase space close to
any collapse-dispersal separatrix leads generically to the prediction
of power-law behavior for the mass of the resulting condensed object.
In order to make this point more explicit, let us consider some set of
initial data parameterized by the control parameter $A$. Let us also
assume that for $A < A_\mathrm{critical}$ the cloud completely
disperses, while for $A > A_\mathrm{critical}$ at least part of the
cloud condenses. In other words, if $A$ lies inside of the
intermediate collapse basin, the solution to the field equations will
be equivalent to finding an integral curve in phase space from $A$ to
the final attractor of this basin. Similarly, if $A$ lies inside of
the dispersal basin, then the integral curve will start at $A$ but end
at the final attractor of dispersal.  Then, under the mild technical
assumption of the existence of at least one critical collapse solution
on the critical surface with exactly one unstable
mode~\citep{Harada03,Harada01,Gundlach1,Gundlach2,Gundlach3,Gundlach4},
the condensed mass will be given by
\be 
\label{scaling}
M \approx M_0 \; [A-A_\mathrm{critical}]^\delta, 
\ee 
provided that the initial data is chosen reasonably close to the
critical surface, {\it{i.e.}} $A \approx A_\mathrm{critical}$.

The physically interesting quantity is the exponent $\delta$, which
arises naturally as the fractional power-series exponent of a
generalized Frobenius expansion for linear perturbations around the
critical solution~\citep{Yunes}.  The order parameter $A$, the value
of the critical initial data $A_\mathrm{critical}$, and the constant
of proportionality $M_0$ can be changed at will by reparameterizing
the initial data set. In contrast, the exponent $\delta$ is physically
significant and reparameterization invariant, depending only on the
equation of state and the condensation mechanism. Observe that, by
construction, we must have $\delta >0$, since $\delta<0$ would imply
an abrupt transition between no condensation and complete condensation
of the cloud. Even $\delta=0$ is problematic, since this corresponds
to an abrupt transition from no condensation to a finite condensate
mass. It is only for $\delta>0$ that as we fine-tune the control
parameter $A$ we get the physically reasonable situation of no
condensation connected smoothly to a low mass condensate for $A >
A_\mathrm{critical}$.

Behavior of this type has by now been seen in numerical experiments in
a considerable number of seemingly disparate situations.
\begin{itemize}
  
\item In Newtonian gravity coupled to a gas cloud with some specified
  equation of state, such as an isothermal one, it is possible to
  observe the same scaling behavior of the mass. First, one searches
  for solutions describing critical collapse and then linearizes
  around these critical collapse solutions to find
  $\delta$~\citep{Harada03,Harada01}. The Newtonian isothermal
  collapse case is of particular relevance in astrophysics, since it
  is a good description for cold molecular gas in the interstellar
  medium, where the cooling time is much shorter than the dynamical
  time.
  
\item In general relativity, the special case where the condensed
  object is a black hole is known as Choptuik
  scaling~\citep{Gundlach1,Gundlach2,Gundlach3,Gundlach4,Choptuik,%
    Choptuik96}.  This phenomenon has now been analyzed not just for
  gas clouds but also for several other forms of matter. In
  particular, the analyses in \citet{Choptuik98} and
    \citet{koikehara} showed that for an adiabatic perfect fluid with
  adiabatic index in the domain $\gamma \in \left(1, 1.89\right)$,
  where $p=(\gamma-1)\rho c^2$, the critical exponent varies over the
  range $\delta \in \left(0.106, 0.817\right)$, clearly demonstrating
  the dependence of this exponent on the equation of state.
\end{itemize}
Several key results are summarized in Table I.

\bigskip

\begin{table}[htb]
\begin{center}
Critical exponents determined by numerical experiment.

\medskip
\begin{tabular}{||l|l||l|l||}
\hline
\hline
System & Critical Point & Exponent $\delta$ & $1/\delta$ \\
\hline\hline
Newtonian isothermal & Hunter A &  0.10567  & 9.4637\\
\hline
GR dust: $p=0$ & Evans--Coleman & 0.10567 &  9.4637   \\
GR radiation: $p=\frac{1}{3}\rho c^2$ & Evans--Coleman & 0.3558019 & 2.810553 \\
GR semi-stiff: $p=\frac{4}{5}\rho c^2$ & Evans--Coleman & 0.73 & 1.37 \\
GR stiff: $p=\rho c^2$ &  Evans--Coleman & 0.96 &   1.04\\
\hline\hline
\end{tabular}
\\[10pt]
{Table I: Key known values of critical exponents in various systems.
  \\ See
  references~\citep{Harada03,Harada01,Gundlach1,Gundlach2,Gundlach3,%
Gundlach4,Choptuik98,koikehara} and references therein.}
\end{center}
\end{table}

Although the phenomenon of Choptuik scaling is concerned with black
hole formation, the qualitative features of the collapse mechanism are
identical to that of stellar formation. Matter subjected to gravity
will generically condense, at least partially, irrespective of whether
it is Einstein's or Newton's gravity that rules. Furthermore, this
condensation will generically occur regardless of the initial form or
distribution of matter.  The collapsing matter knows nothing about its
final fate, or which basin of attraction it lies in.  Hence, the
presence or absence of critical mass scaling will be independent of
the final fate of the system.

In this manner, it is clear that the details of the separatrix that is
being considered are irrelevant, as far as the existence and numerical
value of mass scaling critical exponents are concerned. The precise
values of these exponents will depend on the collapse mechanism and
equation of state, and not on the final fate of the cloud. When we
confront our theoretical ideas with observation, we will use the
numerical exponents found for black hole formation [see
\citep{Harada03,Harada01,Gundlach1,Gundlach2,%
  Gundlach3,Gundlach4,Choptuik98,koikehara} and references therein],
to attempt to pin down the nature of the specific critical collapse
process that actually leads to the formation of subsolar stellar
objects.

\section{From critical collapse to IMF}

Extending this analysis further, suppose a number of Newtonian or
relativistic systems, with initial data depending on some control
parameter $A$, evolve dynamically.  Let the distribution of initial
control parameters be given by the probability distribution function
$P_a(A)$.  We can then determine the probability $P(M)\propto
\mathrm{d}N / \mathrm{d}M $ of producing low-mass condensed objects by
calculating
\be
P(M) \; \d M = P_a(A) \; {\frac{\d A}{\d M}} \; \d M. 
\ee
We can invert equation~(\ref{scaling}), to obtain
\be
A(M) \approx A_{\mathrm{critical}} + \left({M\over M_0}\right)^{1/\delta},
\ee
and so rewrite the probability as
\be
P(M) \; \d M \approx {\frac{1}{\delta}} \; {\frac{P_a(A)}{M_0}} \;  
\left(\frac{M}{M_0}\right)^{(1/\delta)-1} \; {\d M}.
\ee
This can further be written as
\be
P(M) \; \d M \approx {\frac{1}{\delta}} \; 
{\frac{P_a(A_\mathrm{critical} + [M/M_0]^{1/\delta})}{M_0}} \;  
\left(\frac{M}{M_0}\right)^{(1/\delta)-1} \; {\d M}.
\ee
As long as $P_a(A)$ is smooth enough to have a Taylor series expansion
we can write
\be
P(M) \; \d M \approx {\frac{1}{\delta}} \; 
\left\{
P_a(A_\mathrm{critical}) \;  
\left(\frac{M}{M_0}\right)^{(1/\delta)-1} 
+
\left.{\d P_a\over\d A}\right|_{A_\mathrm{critical}} 
\left(\frac{M}{M_0}\right)^{(2/\delta)-1} 
\right\}
\; \frac{\d M}{M_0}.
\ee
As long as $\delta$ is positive, there will be some region for which
the linear term and all higher-order terms can safely be neglected,
since $M$ is assumed small as compared to $M_0$. We can then rewrite
the probability as
\be 
P(M) \; \d M \approx {\frac{1}{\delta}} \;
P_a(A_\mathrm{critical}) \; 
\left(\frac{M}{M_0}\right)^{(1/\delta)-1} \; \frac{\d M}{M_0},
\qquad \hbox{as} \quad  M \ll M_0.  
\ee 
This restriction that $P_a(A)$ have a well-behaved Taylor expansion
near the critical value of the order parameter is not strong at all.
With this in mind, we expect for low mass objects a power law
distribution in masses:
\be
P(M\ll M_0) \approx 
{\frac{\cal A}{M_0}} \; \left({\frac{M}{M_0}}\right)^{(1/\delta)-1}
=
{\frac{\cal A}{M_0}} \; \left({\frac{M}{M_0}}\right)^{n-1},
\ee
where we have defined $n \equiv 1/\delta$.  We must emphasize that the
previous equation is completely equivalent to either of the following
forms:
\be
P(M\ll M_0) \approx 
{\frac{1}{M_0}} \; \left({\frac{M}{M_\delta}}\right)^{(1/\delta)-1}
=
{\frac{1}{M_0}} \; \left({\frac{M}{M_\delta}}\right)^{n-1}
.
\ee
or
\be
P(M\ll M_0) \approx 
{\frac{1}{\widetilde M_\delta}} \; 
\left({\frac{M}{\widetilde M_\delta}}\right)^{(1/\delta)-1}
=
{\frac{1}{\widetilde M_\delta}} \; 
\left({\frac{M}{\widetilde M_\delta}}\right)^{n-1}
.
\ee
Any of these three forms differ from each other only by a redefinition
of the parameters appearing in the power law. This implies an extreme
flexibility in the appearance of the power law, which may sometimes
disguise the equality of two apparently different presentations. The
only true invariant of a power law is the exponent $\delta$ (or $n$).
The mass parameters $M_0$, $M_\delta$, and $\widetilde M_\delta$, can
be redefined almost at will.

Furthermore, observe that this entire analysis holds only for small
masses, since we have assumed that the control parameter $A$ is near
the critical surface.  This behavior is structurally similar to the
observed high-mass IMF, $\xi(M)$, given by
\be 
\xi(M) = {\d N\over\d M} = N_\mathrm{total} \; P(M),
\ee
where $N_\mathrm{total}$ is the total number of stars in the region of
interest and at large mass the probability function is a power law of
the form
\be 
P(M\gg M_0) \approx {\frac{\cal B}{M_0}} \; 
\left(\frac{M}{M_0}\right)^{-m-1}. 
\ee
At large mass observation favors the Salpeter exponent $m\approx
1.35$. [Unfortunately not all authors agree on the precise definition
of $\xi(M)$, see \citet{Miller-Scalo} for a brief discussion, and
accordingly some care must be taken in comparing IMFs extracted from
different sources.] We choose to work directly with the probability
distribution $P(M)$ normalized so that
\be 
\int_0^\infty P(M) \; \d M = 1.
\ee

The major difference at low mass is that the sign of the exponent
changes, which is necessary on two counts: in order that the
probability function be integrable, and that the exponent $\delta$ be
even in principle calculable within the current scenario.

\section{Modeling the IMF}

The simplest toy model that exhibits both forms of asymptotic behavior
and that is similar to models widely used by astronomers to 
describe the observations is
\be
P(M) =  {\frac{n \; m}{n+m}} \; {1\over M_0} \; 
\left\{
\left(\frac{M}{M_0}\right)^{+n-1}  \Theta(M_0-M)
+
\left(\frac{M}{M_0}\right)^{-m-1}  \Theta(M-M_0)
\right\},
\ee
where both $n$ and $m$ are positive. This toy model provides us with a
well-behaved normalizable probability distribution
\be
\int P(M)\; \d M = 1; \qquad (m>0; \;\; n>0);
\ee
with finite mean and variance:
\be
\overline M = \int M\; P(M)\; \d M = {nm\over (n+1)(m-1)} M_0;
\qquad (m>1; \;\; n>0)
\ee
\be
\sigma^2 = \int (M-\overline M)^2\; P(M)\; \d M  =
{ nm[ (m-1)^2 + (n-1)^2 - 1] \over (n+2)(n+1)^2(m-2)(m-1)^2} \; M_0^2;
\qquad (m>2; \;\; n>0).
\ee
This is the simplest model that is realistic in terms of being well
behaved at both the upper and lower limits. 

More generally, one might wish to consider piecewise power laws as
suggested by~\citet{Miller-Scalo,Scalo,Kroupa}. Note that interest in
these piecewise power laws is dictated by their mathematical and
observational convenience --- there is no sound physical motivation
for the abrupt change in behaviour encoded in the Heaviside function.
However, this distribution proves effective in summarizing the
observational data with a minimum of complications.  A generalized
representation of these piecewise power laws is
\be
P(M) = \sum_{i=0}^N {{\cal B}_i \over \mu} \;\;
\theta(M-M_i) \;\; \theta(M_{i+1}-M)\; \;
\left(\frac{M}{\mu}\right)^{n_i-1},
\ee
with $M_0 = 0$, and $M_N=\infty$. Here $\mu$ is any conveniently
chosen arbitrary but fixed mass scale, and the coefficients ${\cal
  B}_i$ are chosen so as to make $P(M)$ continuous:
\be
{\cal B}_i \left(\frac{M_{i+1}}{\mu}\right)^{n_i-1} =
{\cal B}_{i+1} \left(\frac{M_{i+1}}{\mu}\right)^{n_{i+1}-1}
\ee
That is
\be
{\cal B}_i =
{\cal B}_{i+1} \left(\frac{M_{i+1}}{\mu}\right)^{n_{i+1}-n_i}
\ee
Observe that although this probability function is continuous, it is
not differentiable at the interfaces $M=M_{i}$, leading to ``kinks''.
This non-differentiability is due to the fact that there is only one
degree of freedom available in the ${\cal B}_i$ and it has been
already used to impose continuity.

In order for this piecewise power law distribution to be normalizable
we must demand $n_0>0$, and $n_N<0$. To additionally obtain a finite
mean we need $n_0>0$, and $n_N<-1$. For a finite variance we need
$n_0>0$ and $n_N <-2$. Thus the values of both the low-mass $n_0$
exponent and high mass $n_N$ exponent have important implications for
the mathematical existence of a normalizable probability distribution,
with finite mean and standard deviation. We shall confront these
theoretical considerations with the observational data in a later
section.

First, however, let us consider the possibility that there are several
\emph{different} and \emph{independent} competing collapse processes
with different critical solutions, indexed by the label
$\alpha$. Then, for \emph{each} of these processes we will have an
independent mass scaling law of the form
\be
M \approx M_{\delta_\alpha} \; (A-A_\mathrm{critical})^{\delta_{\alpha}}.
\ee
If the condensation is known to take place via process $\alpha$,
then this leads to a probability distribution given by
\be
P_{\alpha}(M) \approx 
{\frac{{\cal A}_\alpha}{M_{\delta_\alpha}}}
\left(\frac{M}{M_{\delta_\alpha}}\right)^{(1/\delta_\alpha)-1}.
\ee
Since none of these competing processes depend on each other, each
will contribute independently with probability $p_\alpha$ to the
total probability distribution, leading to
\be
\label{kinked}
P_\mathrm{total}(M) \approx 
\sum_{\alpha} p_\alpha \; {P_{\alpha}(M)} = 
\sum_\alpha p_\alpha \; {\frac{{\cal A}_\alpha}{M_{\delta_\alpha}}} \; 
\left(\frac{M}{M_{\delta_\alpha}}\right)^{(1/\delta_\alpha)-1}.
\ee
By redefining parameters we can rewrite this to provide
an alternative but equivalent representation of the form
\be
\label{kinked2}
P_\mathrm{total}(M) \approx  
\sum_\alpha \; {\frac{{\cal B}_\alpha}{\mu}} \; 
\left(\frac{M}{\mu}\right)^{(1/\delta_\alpha)-1}
=
\sum_\alpha \; {\frac{{\cal B}_\alpha}{\mu}} \; 
\left(\frac{M}{\mu}\right)^{n_\alpha-1}.
\ee
We emphasize that this is a sum over \emph{distinct} and
\emph{independent} critical solutions, and so is not intrinsically a
``piecewise power law'' of the type considered above.  Although this
physically motivated distribution possesses the same general shape as
the ``kinked'' probability law aforementioned, it smoothes out the
``kinks'', allowing for differentiability at the interfaces.  The
general shape of the ``kinked'' power law is retained because, the
largest of the $\delta_\alpha$ will dominate at the smallest masses.
Eventually, there will be an approximate switch-over to one of the
other critical exponents at larger mass.  If this larger mass is still
reasonably small, one could calculate it using dynamical system
theory. In this manner, one may hope to approximately model the
observationally-motivated piecewise power-law IMF all the way up to
its peak with a physically-motivated differentiable power-law.
However, one can never obtain the high-mass decreasing tail from this
sort of analysis.

Finally, we should mention the possibility of using Gamma and
inverse-Gamma distributions as building blocks for the IMF probability
function. These distributions require additional physical input in the
form of an exponential cutoff, which, although not well motivated
physically, allows for the construction of global fits to the data.
For the Gamma distribution
\be
P_n(M) = \frac{1}{\Gamma(n)} \; 
\left(\frac{M}{M_0}\right)^{n-1} \; 
\exp\left\{ - \frac{M}{M_0} \right\}.
\ee
The probability integral converges for $n>0$, and has finite mean and
variance
\be
\overline M = \int M\; P(M)\; \d M = n \;M_0;
\ee
\be
\sigma^2 = \int (M-\overline M)^2\; P(M)\; \d M  =
n \; M_0^2.
\ee
This Gamma distribution is most useful for low mass where it provides
an accurate approximation to a pure power law, while at high mass the
exponential cutoff keeps everything finite. In contrast, the
inverse-Gamma distribution is
\be
P_m(M) = \frac{1}{\Gamma(m)} \; 
\left(\frac{M}{M_0}\right)^{-m-1} \; 
\exp\left\{ - \frac{M_0}{M} \right\}.
\ee
The probability integral now converges for $m>0$, and has finite mean
and variance
\be
\overline M = \int M\; P(M)\; \d M = \frac{M_0}{m-1}; \qquad (m>1)
\ee
\be
\sigma^2 = \int (M-\overline M)^2\; P(M)\; \d M  =
\frac{M_0^2}{(m-2)(m-1)^2}; \qquad (m>2).
\ee
The inverse-Gamma distribution is most useful for high mass where it
provides an accurate approximation to a pure power law, while at low
mass the exponential cutoff keeps the total probability finite.

Arbitrarily complicated probability distributions could now be
constructed by taking linear combinations of Gamma and inverse-Gamma
distributions. This is, however, not the way things have historically
been done in the observational literature. These comments on Gamma
distributions are mentioned here as a potentially useful representation
for future study in observational astrophysics.

\section{Observational situation}

In contrast to these theoretical considerations, direct astrophysical
observation leads to several models for $P(M)$ that are piecewise
power laws (Table II), and to several isolated data points at low mass
(Table III). The three standard IMFs are those of
Salpeter~\citep{Salpeter}, Miller--Scalo~\citep{Miller-Scalo}, and
Scalo~\citep{Scalo}, with a more recent version due to
Kroupa~\citep{Kroupa}. Relatively few of the ranges in Table II
correspond to a positive $\delta$. For low mass condensates, Scalo
gives $m = -1/\delta= -2.60$ so that $\delta =0.385$, while Kroupa
gives $m = -1/\delta \in ( -1.4, 0.0) $ so that $\delta \in ( 0.71,
\infty)$. All the other parts of the standard IMFs correspond to the
high mass region where the number density is decreasing with
increasing mass.

\begin{table}[htb]
\begin{center}
Multi-scale observational IMFs.

\medskip
\begin{tabular}{||l|c|c||c||}
\hline
\hline
IMF: $P(M) = ({\cal A}/M_0) \; (M/M_0)^{-m-1}$ &
 $M_1/M_\odot$ & $M_2/M_\odot$ & Exponent $m$\\
\hline\hline
\citet{Salpeter} & 0.10  & 125 & 1.35\\
\hline\hline
\citet{Miller-Scalo} & 0.10& 1.00 & 0.25\\
 & 1.00& 2.00& 1.00\\
 & 2.00& 10.0& 1.30\\
 & 10.0& 125& 2.30\\
\hline\hline
\citet{Scalo} & 0.10& 0.18 & $-2.60$\\
 & 0.18 & 0.42 & 0.01\\
 & 0.42& 0.62 & 1.75\\
 & 0.62& 1.18 & 1.08\\
 & 1.18 &3.50 & 2.50\\
 & 3.50& 125& 1.63\\
\hline\hline
\citet{Kroupa} & 0.01 & 0.08 & $-0.7\pm 0.7$ \\
 & 0.08 & 0.50 & $+0.3\pm0.5$ \\
 & 0.50 & $\infty$ & $1.3\pm0.3$ \\
\hline\hline
\end{tabular}
\\[10pt]
{Table II: Observationally derived piecewise power-law $P(M)$.}
\end{center}
\end{table}

Those IMFs obtained using observations which focused on the substellar
regime are summarized in Table III. These observations indicate broad
observational agreement as to the sign of the low-mass exponent, and a
preponderance of evidence pointing to a clustering of the exponent at
$m\approx -0.5$, {\it{i.e.}} $n\approx +0.5$ and $\delta \approx +2$.
These low-mass exponents are converted into critical exponents in
Table IV.  By comparing the theoretical results in Table I with the
observational results in Table IV, we can see that while there is
broad agreement between observation and theory regarding the sign of
the exponent, \emph{quantitative} agreement is more problematic.

\begin{table}[htb]
\begin{center}
Low-mass observational IMF.

\medskip
\begin{tabular}{||l|c|c||c||}
\hline\hline
IMF: $P(M) = ({\cal A}/M_0) \; (M/M_0)^{-m-1}$ &
 $M_1/M_\odot$ & $M_2/M_\odot$ & Exponent $m$\\
\hline\hline
\citet{Barrado00} &0.2& 0.8 & $-0.2$\\
\citet{Barrado} &0.035& 0.3 & $-0.4$\\
\citet{Bouvier,Bouvier2}  & 0.03 & 0.48 & $-0.4$\\
\citet{Martin} &0.02&0.1  & $-0.47$\\
\citet{Bouvier0209}  & 0.072 & 0.4 & $-0.5$ \\
\citet{Luhman} & 0.02 & 0.1  & $-0.5$ \\
\citet{Najita} & 0.015 & 0.7 & $-0.5$\\
\citet{Rice} 
& $10^{-5}$ & $10^{-3}$ & $\approx -1$\\
\hline
\citet{Tej} & 0.01 & 0.50 & $-0.2\pm 0.2$\\
\citet{Tej} & 0.01 & 0.50 & $-0.5\pm 0.2$ \\
\hline\hline
\end{tabular}
\\[10pt]
{Table III: Observationally derived low-mass $P(M)$.}
\end{center}
\end{table}

\bigskip

\begin{table}[htb]
\begin{center}
Observed low-mass exponents.

\medskip
\begin{tabular}{||l|l||l|l||}
\hline
\hline
Source & Exponent $m$ & Exponent $1/\delta$ &  Exponent $\delta$ \\
\hline\hline
\citet{Scalo} & $-2.60$ & 2.60 & 0.385 
\\
\citet{Kroupa} & $-1.4$ --- $0.0$ &$0.0$ --- $1.4$ & $0.71$ --- $\infty$
\\
\hline
\citet{Rice}  &$\approx -1$ & $\approx 1 $ & $\approx 1$
\\
\citet{Najita} & $-0.5$ & 0.5 & 2.0 
\\
\citet{Luhman} & $-0.5$ & 0.5 & 2.0 
\\
\citet{Bouvier0209}  &  $-0.5$ & $0.5$ & $2.0$
\\
\citet{Martin} & $-0.47$ & $0.47$ & $2.16$
\\
\citet{Bouvier,Bouvier2}  &  $-0.4$ & $0.4$ & $2.5$
\\
\citet{Barrado} & $-0.4$ & 0.4 & 2.5
\\
\citet{Barrado00} & $-0.2$ & 0.2 & 5.0
\\
\hline
\citet{Tej} & $-0.5$ & 0.5 & 2.0\\
\citet{Tej}& $-0.2$ & 0.2 & 5.0 
\\
\hline\hline
\end{tabular}
\\[10pt]
{Table IV: Observational estimates of the very low mass exponents.}
\end{center}
\end{table}

A particularly nice feature is that the observationally derived low
mass exponent $n_0=-m_0\approx 1/2$ is compatible with a normalizable
probability distribution. The observationally determined high mass
exponent (the Salpeter exponent) $m_N=-n_M\approx 1.35$ is compatible
with a normalizable probability distribution of finite mean, but with
an infinite variance arising from the high-mass tail. Probability
distributions of finite mean but infinite variance are well-known in
statistics, and while they make perfectly good sense mathematically
they are associated with perhaps unexpected mathematical subtleties
(such as, for instance, the failure of the central limit theorem).
Note that this behavior is coming from the high-mass region, not the
low-mass region that is of primary concern in the current article.

We must conclude that present day observational data is sufficiently
poor that the only rigorous inference one can draw is that the
exponent has changed sign at sufficiently low masses. Beyond that, it
would be desirable to contrast the exponent occurring in the subsolar
IMF with the exponent arising in a specific critical collapse process.
Unfortunately, neither observational data nor theory is currently well
enough developed to do so with any degree of reliability.  

Some of the numerical simulations give critical exponents that overlap
with some of the observations. For instance, the Scalo exponent is
roughly comparable with that arising from numerical simulations of
collapse of a relativistic radiation fluid, $p={\frac{1}{3}}\rho c^2$.
Part of the range of Kroupa's IMF, {\it{i.e.}} $\delta\in(0.71,1)$, is
compatible with simulations of a relativistic adiabatic perfect fluid,
$p = k\;\rho c^2$ with $k\in({\frac{4}{5}},1)$. Finally, the IMF
exponent of Rice \emph{et al} is compatible with a numerical critical
solution corresponding to a relativistic stiff fluid, $p = \rho c^2$.
Those observations that cluster around $\delta=2$ are not compatible
with any \emph{known} critical collapse solution. This might indicate
either a problem with the observational data, or a more fundamental
lack of understanding regarding the physically relevant critical
collapse process.

\section{Conclusions}

Future work along these lines should be focused in two directions.
Observationally, improved data would be desirable to test the
hypothesis that the low-mass exponent $\delta$ is both positive and
universal.  Theoretically, it would be important to understand
\emph{quantitatively} why critical behavior provides an accurate
representation of the IMF for $M \lesssim 0.8 ~M_{\odot}$.  It is
clear that as the final condensed mass increases, the initial data $A$
is pushed farther away from the critical surface, {\it{i.e.}} $A \neq
A_\mathrm{critical}$.  Although it is known that the linear
perturbation around the critical solution then loses validity, a
precise calculation of the region of convergence is still lacking.

The formation of real-world gravitational condensates is likely to
involve rotating turbulent dust clouds.  Therefore, it would be very
useful to understand the influence of both angular momentum and
turbulence on the theoretically derived critical exponents. Physical
intuition suggests that turbulence would make the collapse process
even more scale-invariant, since it would eliminate all irregularities
in the collapse and favor self-similar behavior. In this manner, the
scaling law and critical exponent presented for spherical collapse
should not be modified by turbulence, but instead it should be made
more predominant.

A detailed analysis of angular momentum presents new challenges to the
critical behavior framework by sometimes introducing a second
non-spherical growing mode that competes with the usual growing
mode~\citep{gundlach01}.  The presence or absence of a second growing
mode depends on the equation of state, and is known to occur for some
specific polytropic equations of state.  These additional growing
modes possess different eigenvalues from the spherical one, leading to
new scaling exponents. In particular, angular momentum will scale with
a new critical exponent and will possess a new critical parameter in
the initial data. In this manner, the probability distribution becomes
now two-dimensional, depending both on mass and angular momentum.
    
At the present time, it is not clear precisely how the functional
form of the mass scaling law will be modified by the inclusion of
angular momentum.  For main sequence stars, however, physical
intuition suggests that low-angular momentum should be dominant.
Hence, corrections to the mass scaling law based on a spherically
symmetric idealization should be small, though not necessarily
negligible.

Our analysis confirms Larson's intuition that stellar formation at low
mass is related (and perhaps even dominated) by chaotic
dynamics~\citep{Larson}.  In particular, the analysis in terms of
dynamical systems theory can be viewed in terms of deterministic chaos
in gravitational collapse.  We do not, however, need to deal with
fractal structures since limit points and limit cycles seem to be
quite sufficient for generating power-law behavior~\citep{Yunes}.  Our
analysis further supports the idea of a universal slope, dependent
only on the relevant critical collapse solution, but independent of
the initial conditions, and disfavors the astrophysical hypothesis of
an IMF that varies in both time and space.

Summarizing, the dynamical exponents found in Newtonian and general
relativistic gravitational collapse can be used to model and
qualitatively explain a power law version of the IMF valid for small
masses.  For the first time, a concrete application to the numerical
phenomena of critical gravitational collapse has been proposed and
tested against observational data.  We have compared these results to
subsolar IMF data and found them in broad qualitative agreement for
low-mass systems, though quantitative agreement is poor at this stage.
The key point is that gravitational collapse naturally leads to power
law behavior in the low mass regime, with an exponent that is opposite
in sign to the observed high-mass behavior.  This provides a new and
fresh view on power-law behavior with specific astrophysical
applications to dynamic gravitational collapse and the IMF.

\acknowledgments

\underline{Acknowledgments:} We are grateful to Ben Owen and Tom Abel
for insightful comments on gravitational collapse theory and
astrophysical observations of the IMF. We would also like to
acknowledge Carsten Gundlach for clarifying some issues regarding the
dynamical systems framework.

Nicolas Yunes wishes to acknowledge the support of the Center for
Gravitational Physics and Geometry, the Center for Gravitational Wave
Physics and the Physics Department at The Pennsylvania State
University.

The research of Nicolas Yunes was supported partly by the US NSF
grants PHY-0114375 and PHY-0245649, while the research of Matt Visser
was supported by a Marsden Grant administered by the Royal Society of
New Zealand.



\end{document}